\documentclass[english]{ieej-e}
\usepackage[dvips]{graphicx}
\usepackage[varg]{txfonts}

\FIELD{}
\YEAR{}
\NO{}
\title{Measurement and Control of Solenoid Stroke using Its Electrical Characteristics}
\authorlist{%
 \authorentry[akita@is.t.kanazawa-u.ac.jp]{Junichi Akita}{n}{KU}
}
\affiliate[KU]{Kanazawa University \\ Kakuma, Kanazawa, 920-1192 Japan\\
}

\begin{document}

\begin{abstract}
In this paper, we describe the algorithm to measure the stroke of solenoid using the electric characteristics of the solenoid, without mechanical attachments.
We also describe the experimental results of controlling the solenoid stroke at intermediate position.
\end{abstract}


\maketitle

\section{Introduction}

Solenoid is the actuator composed of the coil and the plunger.
Solenoid is used to push or pull the object, and the stroke is controlled in two states; ON and OFF, in usual cases.
There are also some researches to control the solenoid stroke at intermediate positions between ON and OFF positions, by controlling the current of the coil to keep the target position of the plunger, where plunger position is measured by external mechanical sensor\cite{control}.

From the viewpoint of electrical characteristics of the solenoid, its inductance changes according to the plunger position, and we basically measure the plunger position with electrical characteristics of the solenoid\cite{measure1,measure2}.
Thus, we can measure and control the solenoid's plunger position without mechanical attachments.
In this paper, we describe the theoretical and experimental analysis of the solenoid's electrical characteristics, and measuring algorithm of the plunger's position.
We also describe the experimental results of the plunger's position control without mechanical attachments.

\section{Electrical Characteristics of Solenoid and Its Driving Circuit}

\begin{figure}[t]
{\hfill
\includegraphics[width=0.6\columnwidth]{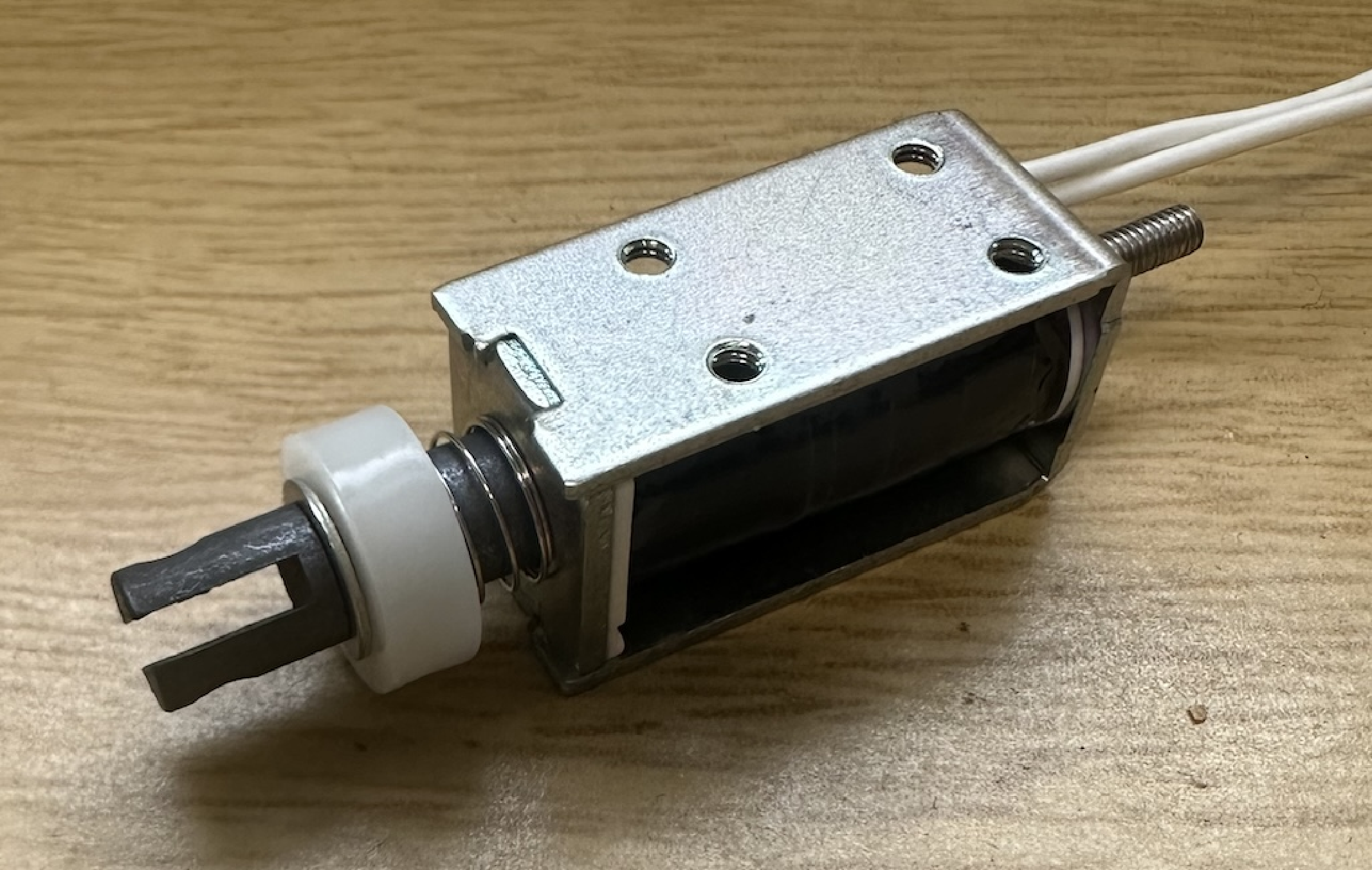}
\hfill}
\caption{Solenoid (Takaha's CBS07300580)}
\end{figure}
\label{fig:solenoid}

\begin{figure}[t]
{\hfill
\includegraphics[width=0.3\columnwidth]{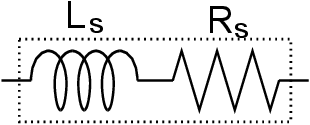}
\hfill}
\caption{Equivalent circuit of solenoid}
\label{fig:solenoid-circuit}
\end{figure}

Solenoid is the actuator composed of the coil and the plunger, as shown in Fig.\ref{fig:solenoid}.
From the viewpoint of electrical characteristics of the solenoid, it is a simple inductor, whose equivalent electrical circuit is described as shown in Fig.\ref{fig:solenoid-circuit}.

\begin{figure}[t]
{\hfill
\includegraphics[width=0.6\columnwidth]{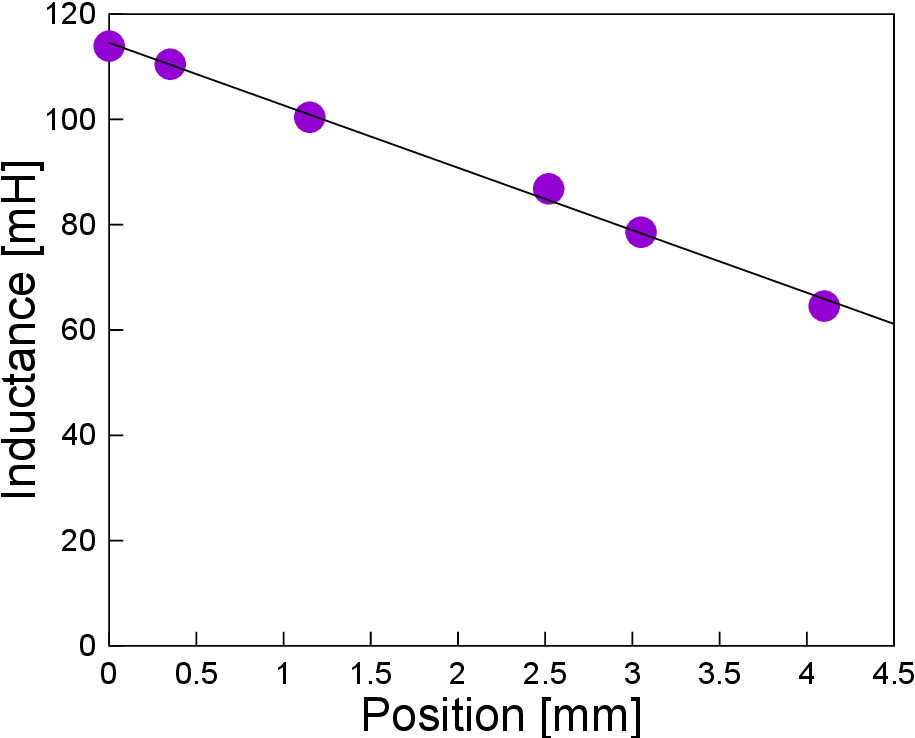}
\hfill}
\caption{Measured induncance and plunger position}
\label{fig:L-pos}
\end{figure}

The plunger in the solenoid is a magnetic material, and it affects the inductance of solenoid, and the position of the plunger also affects the inductance of the solenoid.
Figure \ref{fig:L-pos} shows the measured inductance against the position of the plunger for the solenoid of Takaha's CBS07300580\cite{takahaCBS} at the frequency of 100[Hz], with LCR meter (????).
We can derive the position of the plunger, $x_P$, as follows.
\begin{equation}
  x_P = -0.0843 L + 9.656
  \label{eq:L-pos}
\end{equation}
Thus, we can measure the position of the plunger by measuring the inductance of the solenoid.

\begin{figure}[t]
{\hfill
\includegraphics[width=0.3\columnwidth]{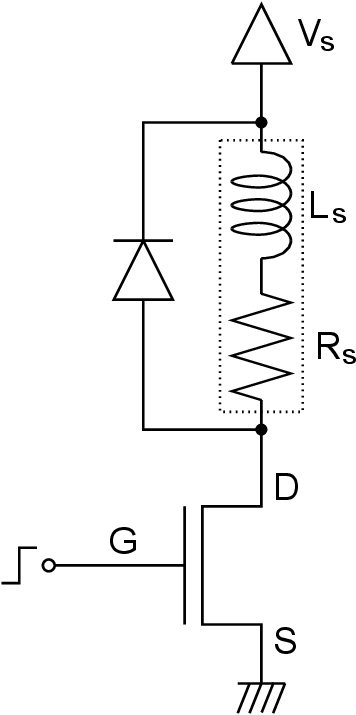}
\hfill}
\caption{Driving circuit of solenoid}
\label{fig:drive-circuit}
\end{figure}

The typical driving circuit of the solenoid is shown in Fig.\ref{fig:drive-circuit}.
Here we define two status of the drive.
\begin{itemize}
\item ON: giving a high voltage to G (gate) to turn the nMOSFET on, where the current flows to make the plunger pulled.
\item OFF: giving a low voltage to G to turn the nMOSFET off, where the current stops to make the plunger released.
\end{itemize}
The transitional phenomenon occurs between switching OFF-to-ON and ON-to-OFF.

\begin{figure}[t]
{\hfill
\includegraphics[width=0.5\columnwidth]{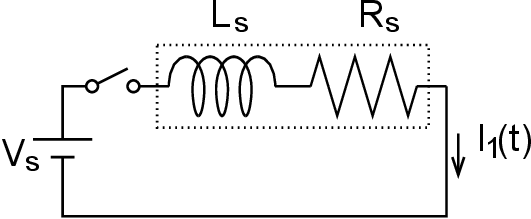}
\hfill}
\caption{Equivalent Circuit for OFF-to-ON transition}
\label{fig:OFF-to-ON}
\end{figure}

At OFF-to-ON transition, the current from the power source gradually increase by the back electromotive force of the inductor, who speed is determined by the time constant of $L_s / R_s$, as shown in Fig.{ref:OFF-to-ON}.
The current of OFF-to-ON transition, $I_1(t)$, the circuit equations is follows.
\[
  L_s \frac{dI_1(t)}{dt} + R_s I_1(t) = V_s
\]
We obtain $I_1(t)$ as follows, with the initial current of $I_1(0) = I_1^S$.
\begin{equation}
  I_1(t) = \frac{V_s}{R_s} - \left( \frac{V_s}{R_s} - I_1^S \right) \exp ( -(R_s/L_s)t )
 \label{eq:I1}
\end{equation}

\begin{figure}[t]
{\hfill
\includegraphics[width=0.4\columnwidth]{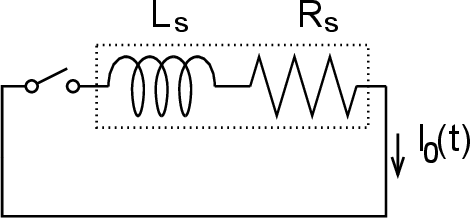}
\hfill}
\caption{Equivalent Circuit for ON-to-OFF transition}
\label{fig:ON-to-OFF}
\end{figure}

At ON-to-OFF transition, the current from the power source stops, and the current gradually decrease by the closed loop by the flywheel diode, as shown in Fig.\ref{fig:ON-to-OFF}.
The current of ON-to-OFF transition, $I_0(t)$, the circuit equations as follows, with assuming the flywheel diode to be a ideal diode, whose equivalent resistance is zero.
\[
  L_s \frac{dI_0(t)}{dt} + R_s I_0(t) = 0
\]
We obtain $I_0(t)$ as follows, with the initial current of $I_0(0) = I_0^S$.
\[
  I_0(t) = I_1^E \exp (-(R_s/L_s)t )
 \label{eq:I0}
\]

\section{Measurement Method of Solenoid Inductance: ON-to-OFF transition}

\begin{figure}[t]
\begin{center}
\includegraphics[width=0.8\columnwidth]{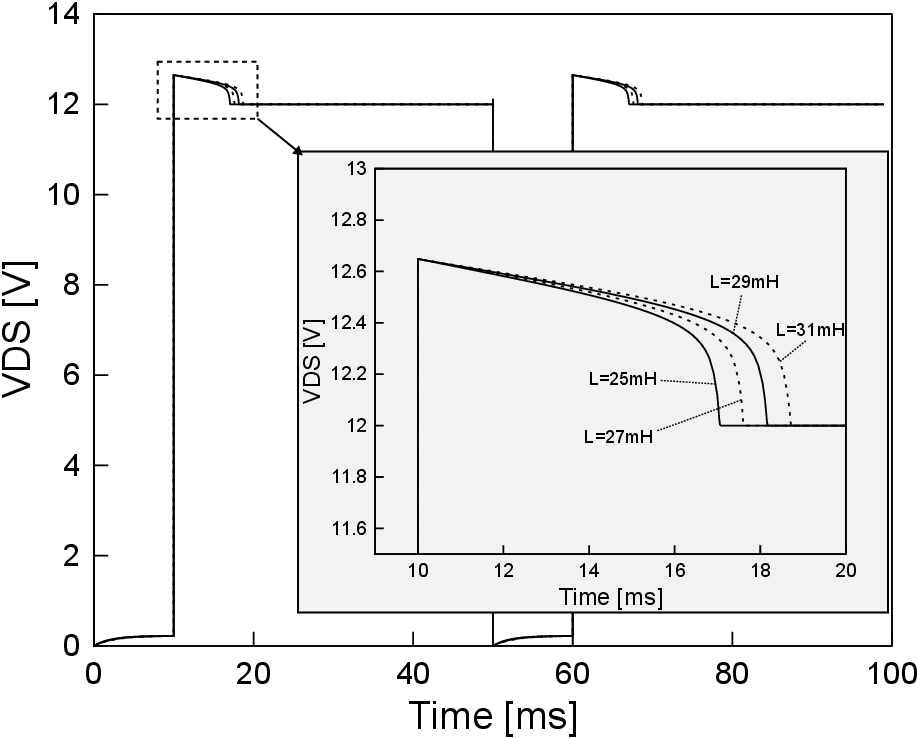}
\end{center}
\caption{Simulated waveforms $V_{\rm DS}$ for ON-to-OFF transition}
\label{fig:sim-ON-to-OFF}
\end{figure}

\begin{figure}[t]
{\hfill
\includegraphics[width=0.9\columnwidth]{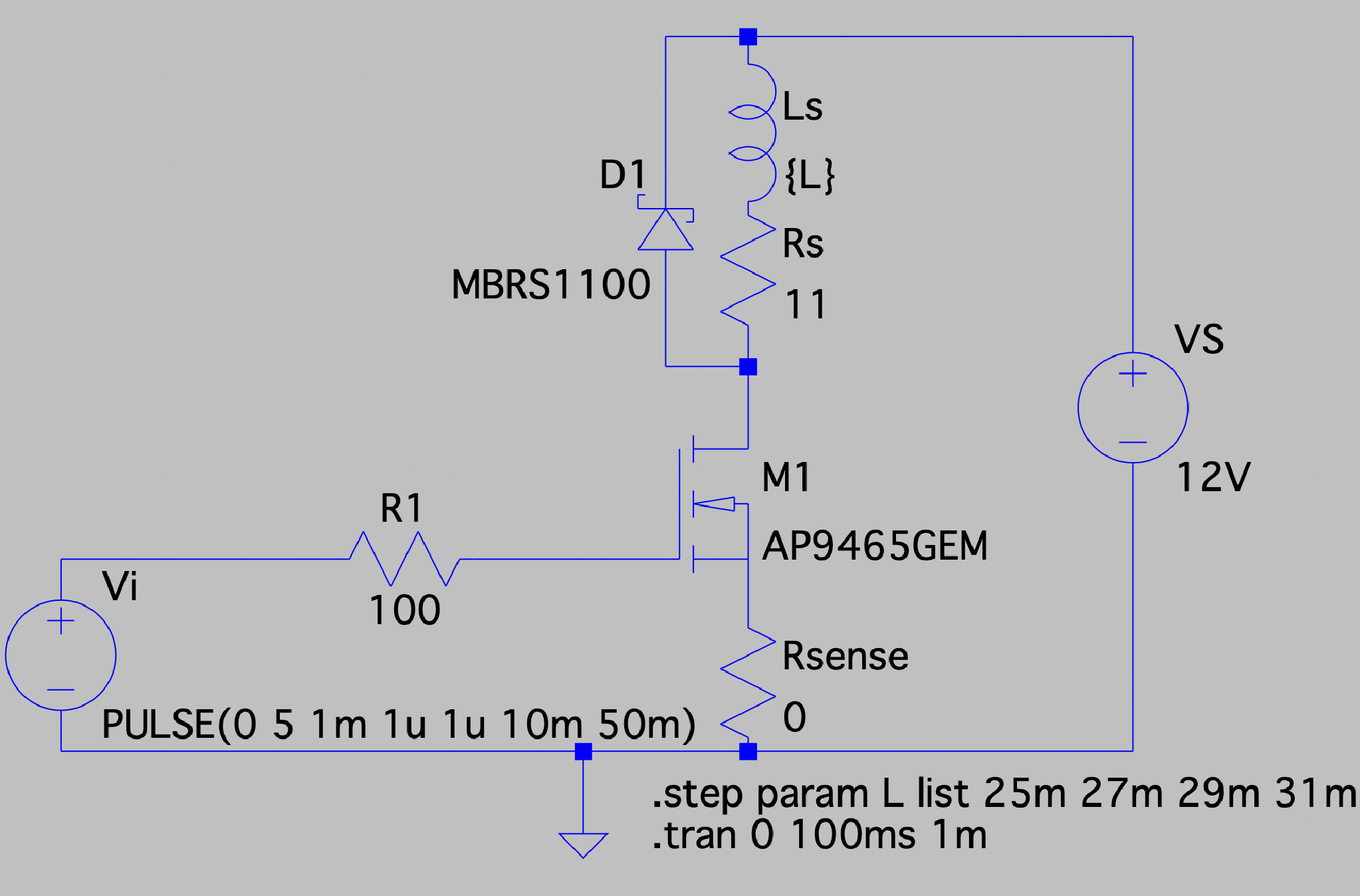}
\hfill}
\caption{Simulation circuit for ON-to-OFF transition}
\label{fig:cir-sim-ON-to-OFF}
\end{figure}

At ON-to-OFF transition, the voltage at D (drain) of the nMOSFET, $V_{\rm DS}$ changes from almost zero (ON reistance of nMOSFET) to $V_s + V_F$, where $V_D$ is the forward voltage of the flywheel diode during the charged energy in the inductor remains.
After the charged energy consumed, the $V_{\rm DS}$ goes $V_s$.
Figure \ref{fig:sim-ON-to-OFF} shows the calculated waveform of $V_{\rm DS}$ by LTspice, for the circuit of Fig.\ref{fig:cir-sim-ON-to-OFF}, with three variations of the inductance.
The time of $V_{\rm DS}$ goes from $V_s + V_F$ to $V_s$ is proportional to the inductance, hence the position of the plunger.
However, this time is 10[ms] at least.
It means that we have to make the solenoid OFF at least 10[ms], which makes the plunger at the released state.
The mechanical time constant of the plunger is defined by the mechanical structure of the solenoid, which is at the order of 10[ms].
It is impossible to keep the plunger at ON or the intermediate position between ON and OFF with having this ``measurement'' period of the ON-to-OFF transition.

\section{Measurement Method of Solenoid Inductance: OFF-to-ON transition}

At OFF-to-ON transition, the current of the nMOSFETf at S (source) or D changes as $I_1(t)$ in Eq.(\ref{eq:I1}).
From the viewpoint of controlling the solenoid stroke, the duty ratio of PWM (pulse width modulation) signal at nMOSFET's G is 0[\%] to 100[\%], and the small duty ratio, such as 5[\%] makes no effect to make the pull operation for the plunger.
We can measure the inductance by the transitional phenomenon at OFF-to-ON transition, with controlling the position of the plunger by controlling PWM duty ratio.

\subsection{Theoretical analysis of PWM drive}

\begin{figure}[t]
{\hfill
\includegraphics[width=0.9\columnwidth]{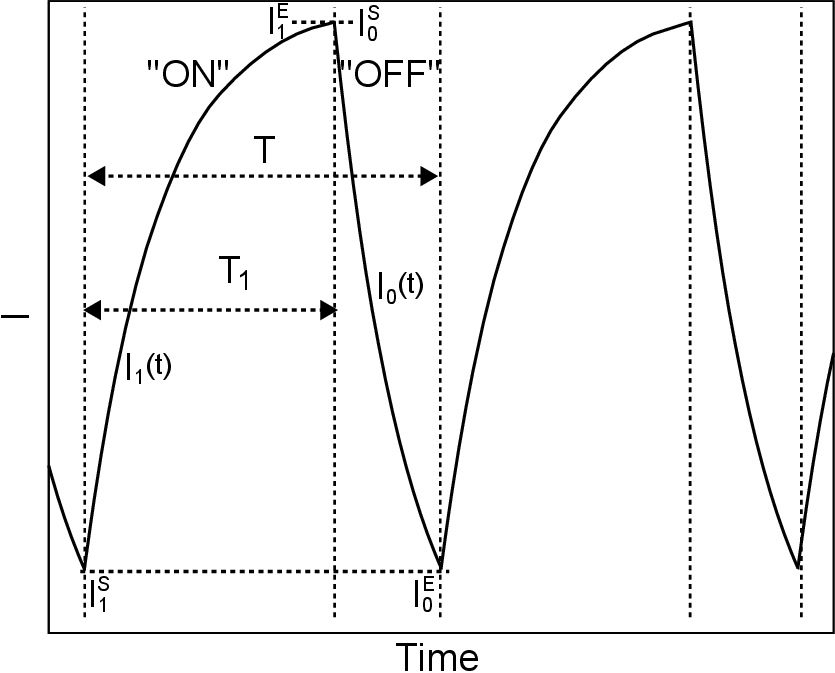}
\hfill}
\caption{Current waveform for PWM drive}
\label{fig:PWMcycle}
\end{figure}

Under PWM control of the solenoid, OFF state starts at the end status of the previous ON state, and next ON state starts at the end status of the previous OFF state, as shown in Fig.\ref{fig:PWMcycle}.
Here we define $I_1^S$ and $I_1^E$ as the initial and the final current during ON state, respectively, and $I_0^S$ and $I_0^E$ as the initial and the final current during OFF state, respectively.
Since the PWM wave is periodical, the initial current of each state equals the final current of previous state, which is expressed as $I_1^S = I_0^E$ and $I_0^S = I_1^E$.
We define the cycle and ON time of PWM wave as $T$ and $T_1$, respectively.
We can calculate the $I_1^S$ and $I_0^S$ as follows from Eqs (\ref{eq:I1}) and (\ref{eq:I0}).
\[
  I_1^E = I_1(T_1) = \frac{V_s}{R_s} - \left( \frac{V_s}{R_s} - I_1^S \right) \exp ( -(R_s/L_s)T_1 )
\]

\[
  I_1^S = I_0(T - T_1) = I_1^E \exp ( -(R_s/L_s)(T - T_1) )
\]
We finally obtain $I_1^E$ and $I_1^S$ as follows, as the function of $R_s$, $L_s$, $V_s$, $T$, and $T_1$.
\[
  I_1^E = \frac{1 - \exp (-(R_s/L_s)T_1)}{1 - \exp (-(R_s/L_s)T)} \frac{V_s}{R_s}
\]
\[
  I_1^S = I_1^E \exp (-(R_s/L_s)(T - T1))
\]

We can also calculate the current at any point during ON (and OFF) state by Eqs (\ref{eq:I1}) and (\ref{eq:I0}).

\subsection{Experimental results and theorical results}
\label{sec:exp}

\begin{figure}[t]
{\hfill
\includegraphics[width=0.9\columnwidth]{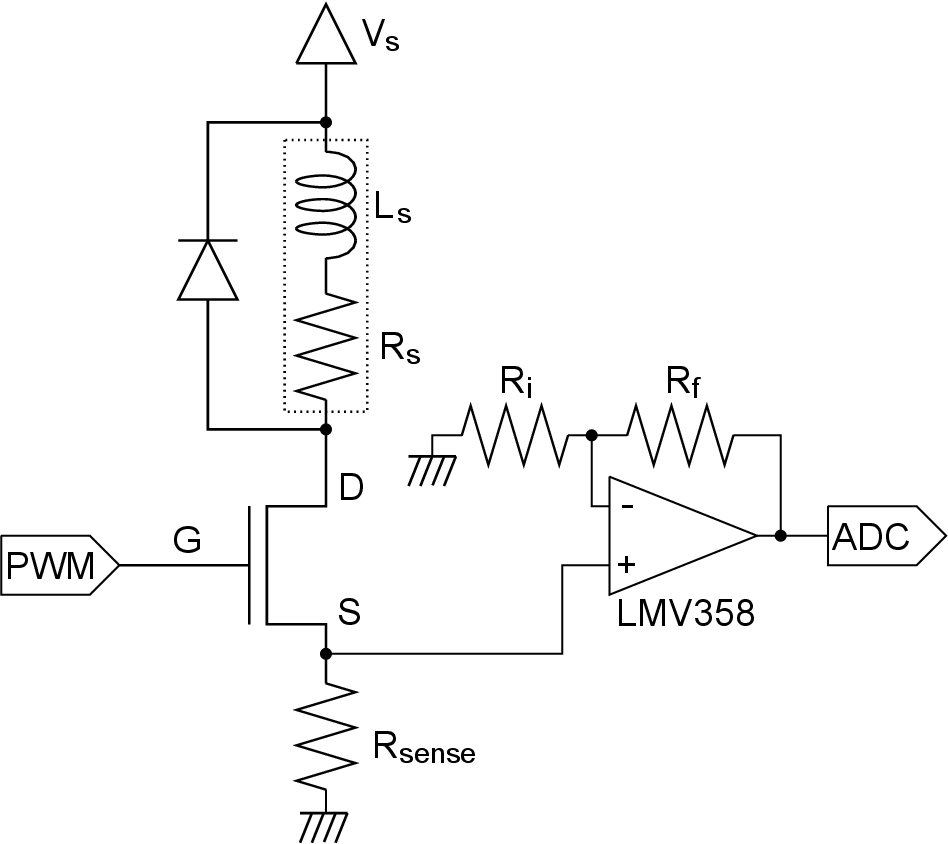}
\hfill}
\caption{Circuit of measurement \& control experiment}
\label{fig:exp-sch}
\end{figure}

We carried out the experimental measurement of $I_1^E$ and $I_1^S$, to find the inductance, $L_s$ or the position of the plunger, from the current and PWM ON time.
Figure \ref{fig:exp-sch} shows the circuit of the experimental setup.
The current of the nMOSFET during ON state is measured as the voltage of $R_{\rm sense} = 0.2[\Omega]$, which is set small so as not to effect the nMOSFET state.
The voltage of $R_{\rm sense}$ is magnified by 101 by the non-inverting amplifier, where $R_i$ and $R_f$ are 1[k$\Omega$] and 100[k$\Omega$], respectively.

We used Arduino UNO \cite{Arduino} for measure and control.
PWM signal is generated using Timer1, and fed to nMOSFET's gate.
$I$ is measured by internal ADC (10bit) every each PWM cycle at 0.1[ms] and 0.5[ms] after the OFF-to-ON transition, by using Timer1's overflow and compare match interrupt.
We set $T = 10[\rm ms]$ (PWM frequency of 100[Hz]), and measured the currents at $T$ of 0.1[ms] and 0.5[ms], $I_{0.1}(T_1, L_s)$ and $I_{0.5}(T_1, L_s)$, respectively, where $T_1$ is ON time of PWM waveform and $L_s$ is the inductance of the solenoid.

\begin{figure}[t]
\begin{center}
\includegraphics[width=0.9\columnwidth]{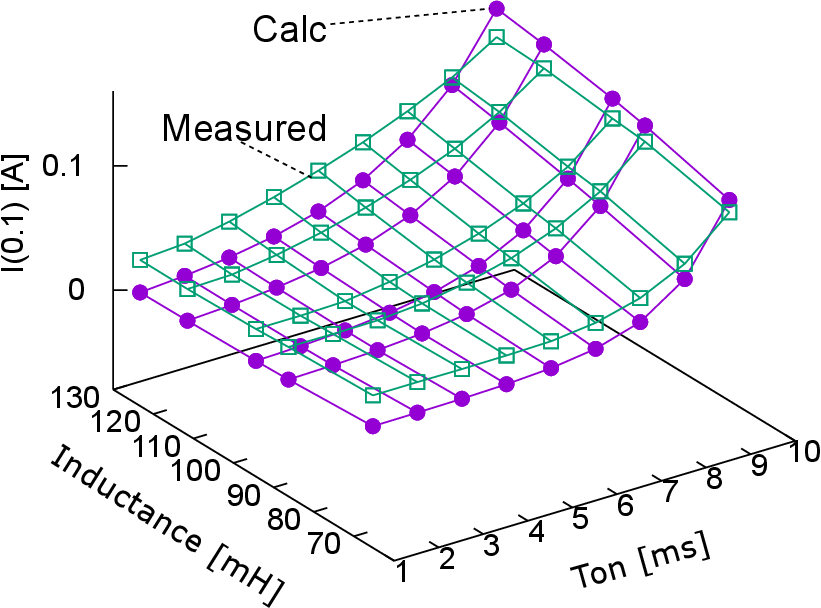}

(a)\vspace*{5mm}

\includegraphics[width=0.9\columnwidth]{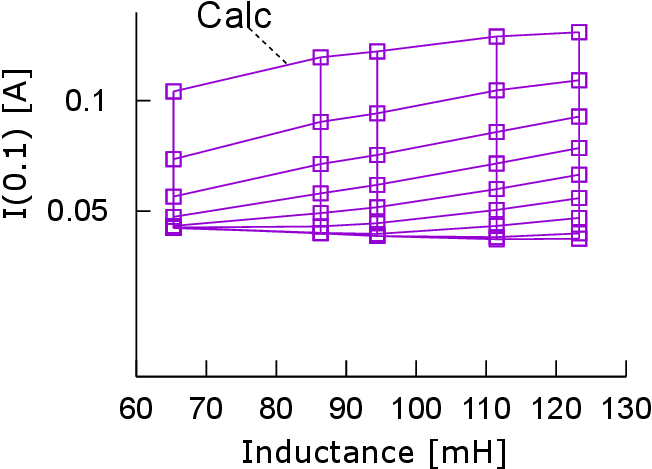}

(b)
\end{center}
\caption{Measured current at 0.1[ms], $I_{0.1}$(a), and its trend against $L$(b)}
\label{fig:exp-res1}
\end{figure}

\begin{figure}[t]
\begin{center}
\includegraphics[width=0.9\columnwidth]{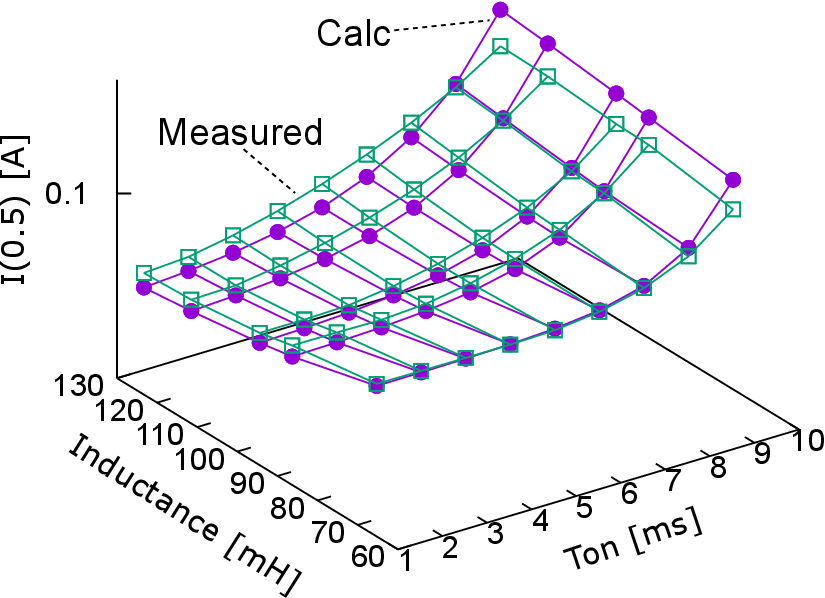}

(a)\vspace*{5mm}

\includegraphics[width=0.9\columnwidth]{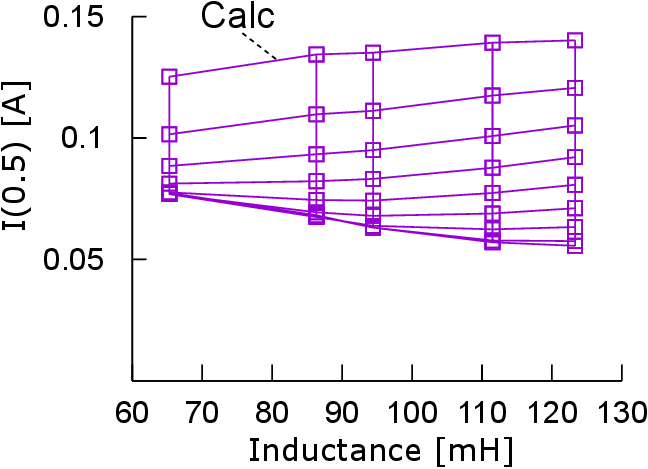}

(b)
\end{center}
\caption{Measured current at 0.1[ms], $I_{0.5}$(a), and its trend against $L$(b)}
\label{fig:exp-res2}
\end{figure}

\begin{figure}[t]
\begin{center}
{\hfill\includegraphics[width=0.9\columnwidth]{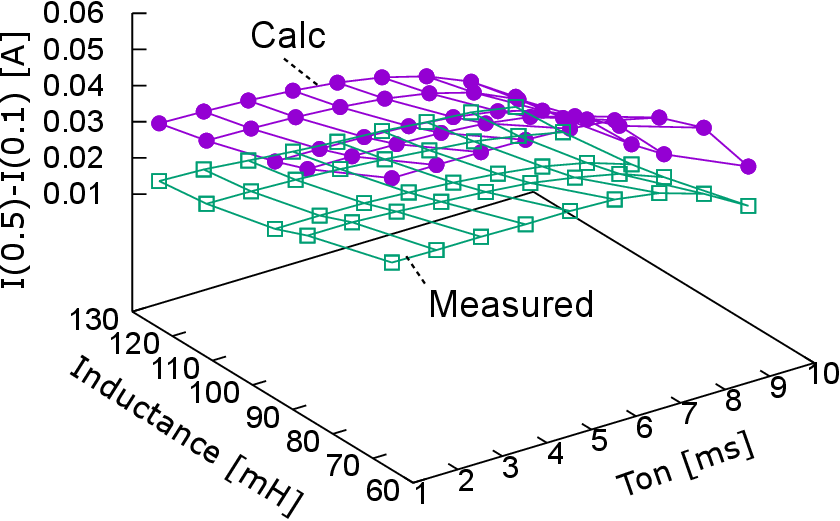}\hfill}
{\hfill (a)\hfill}
\vspace*{5mm}
{\hfill\includegraphics[width=0.9\columnwidth]{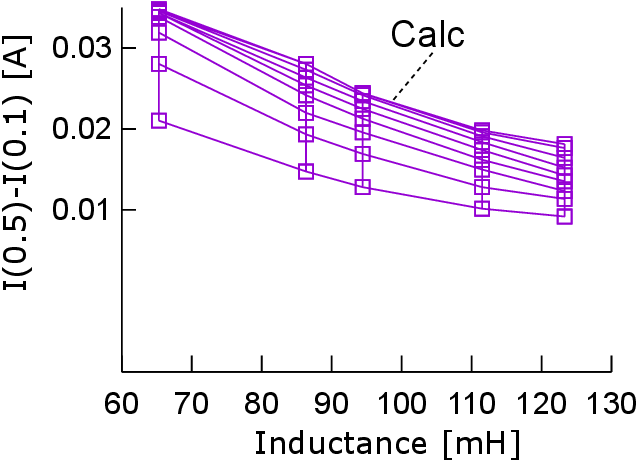}\hfill}
{\hfill (b)\hfill}
\end{center}
\caption{Measured current difference, $I_{0.5} - I_{0.1}$(a), and its trend against $L$(b)}
\label{fig:exp-res3}
\end{figure}

Figure \ref{fig:exp-res1}, \ref{fig:exp-res2} and \ref{fig:exp-res3} shows measured and calculated $I_{0.1}(T_1, L_s)$, $I_{0.5}(T_1, L_s)$, and $I_{0.5}(T_1, L_s) - I_{0.1}(T_1, L_s)$, respectively, for five variations of $L_s$, and $T_1$ from 1[ms] to 9[ms] with step of 1[ms].

Since the calculated currents by Eqs (\ref{eq:I1}) and (\ref{eq:I0}) assumes the ideal flywheel diode, we can see the difference to the measured current, especially for large $T_1$, where the charged energy at the inductor is large.

From Fig.\ref{fig:exp-res1} and Fig.\ref{fig:exp-res2}, the changes of the current at 0.1[ms] and 0.5[ms] over $L_s$ becomes almost flat for some $T_1$.
This means that we can't determine $L_s$ from the measured current at each point.

From Fig.\ref{fig:exp-res3}, the current changes as $L_s$ changes.
This means that we can determine $L_s$ from the measured current at each point, with the control value of $T_1$.

\subsection{Interpolating inductance}

\begin{figure}[t]
{\hfill
\includegraphics[width=0.6\columnwidth]{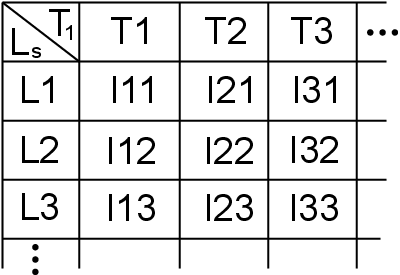}
\hfill}
\caption{Table of current $I$ for $T_{\rm ON}$ and inductance $L_s$}
\label{fig:I-T-L-table}
\end{figure}

Since these phenomena are non-linear, we can't analytically derive the currents for given $L_s$ and $T_1$.
For the practical applications, we can generate the table of the measured current, $I = I_{0.5} - I_{0.1}$ for variations of $L_s$ and $T_1$ in advance.
Here we assume $T_1$ and $L_s$ changes independently, and obtain the measured current for each $T_1$ and $L_s$, $I(T_1, L_s)$, as showin in Fig.\ref{fig:I-T-L-table}.

We assume that we can set $T_1$ as a control parameter.
The purpose of this measurement is to calculate $L_s$ from the measured $I$ and the controlled $T_1$ from $I(T_1, L_s)$.

\begin{figure}[t]
{\hfill
\includegraphics[width=0.5\columnwidth]{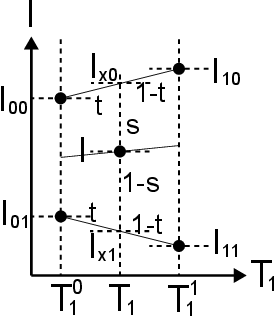}
\hfill}
\caption{Interpolation algorithm}
\label{fig:interpolation}
\end{figure}

Here, set (controlled) $T_1$ exists in the range of $(T_1^0, T_1^1)$, where $T_1^0$ and $T_1^1$ are in the table of $I(T_1, L_s)$.
We can also find the pair of $I$ in the table of $I(T_1, L_s)$ which covers the measured value of $I$.
Thus, we can define the following four points, $(T_1^0, I_{00})$, $(T_1^1, I_{10})$, $(T_1^0, I_{01})$ and $(T_1^1, I_{11})$, which covers the point of $(T_1, I)$, as shown in Fig.\ref{fig:interpolation}.
Each point has the value of $L_s$ from the table of $I(T_1, L_s)$, as described as $L_{00}$, $L_{10}$, $L_{01}$, and $L_{11}$, respectively.
The problem to solve is to calculate $L_s$ from these four points for measured $I$ and controlled $T_1$.
By using the bilinear interpolation, we obtain the interpolation parameter os $t$ and $s$ and the interpolated $I$ in Fig.\ref{fig:interpolation} as follows.
\begin{eqnarray*}
  t &=& \frac{T_1 - T_1^0}{T_1^1 - T_1^0} \\
  I_{x0} &=& I_{00} + t ( I_{10} - I_{00} ) \\
  I_{x1} &=& I_{01} + t ( I_{11} - I_{01} ) \\
  L_{x0} &=& L_{00} + t ( L_{10} - L_{00} ) \\
  L_{x1} &=& L_{01} + t ( L_{11} - L_{01} ) \\
  s &=& \frac{I - I_{x0}}{I_{x1} - I_{x0}}
\end{eqnarray*}
Finally, we obtain the interpolated value of the inductance, $\hat{L}$ from the measured $I$ and the controlled $T_1$ as follows.
\[
  \hat{L} = L_{x0} + s ( L_{x1} - L_{x0} )
\]
We can calculate the position of the plunger, $x_P$, by Eq.(\ref{eq:L-pos}) as follows.
\[
  x_P = -0.0843 \hat{L} + + 9.656 
\]

\section{Control of Solenoid Stroke using Electrical Characteristics}

We carried out the experiment of controlling the solenoid stroke based on the calculated position of the plunger, $x_P$ from the measured $I$ and controlled $T_1$.
We used Arduino UNO as in section {sec:exp}.
The $x_P$ is measured at {\tt loop()} function from the measured $I$, and the PWM duty, next step's $T_1$, $T_1'$ is is calculated as follows based on simple PID control, with the target position of<
\[
  T_1' = T_1 + K_i (x_P - x_T)
\]

\begin{figure}[t]
{\hfill
\includegraphics[width=0.9\columnwidth]{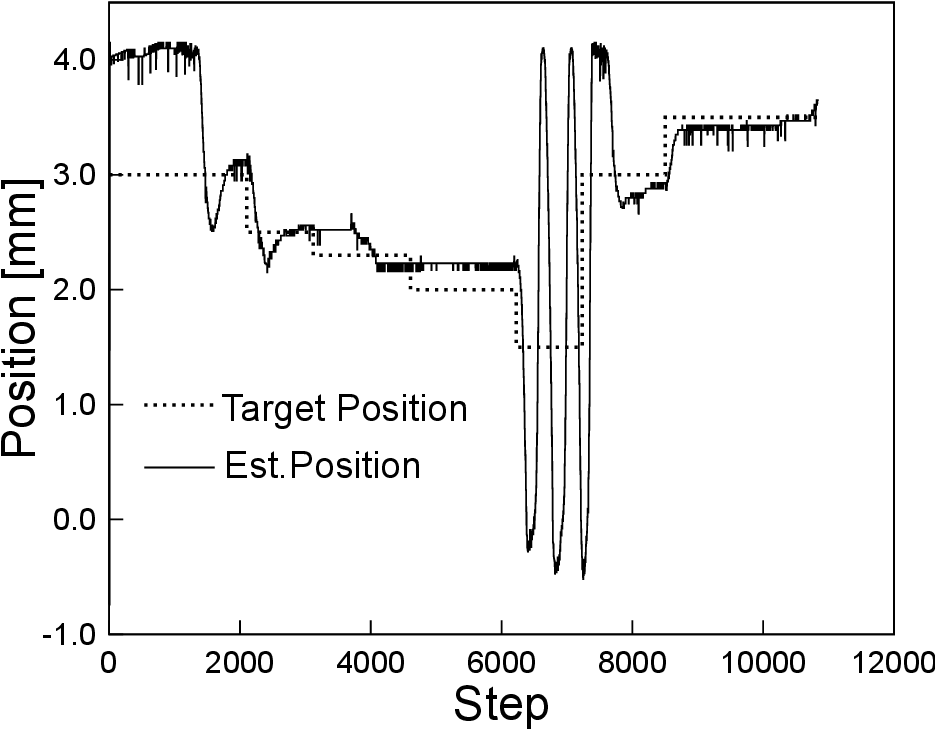}
\hfill}
\caption{Experimental result of solenoid position control}
\label{fig:control-result}
\end{figure}

We set the PID gain of $K_i$ as 4.0 from some experimental trials.
Figure \ref{fig:control-result} shows the measured (calculated) position of the plunger, with the target position of the plunger, where $t$ is the number of steps of control loop.
We can find some overshoots at the target position change of around $t=1800$ and $t=2200$.
We can also find an oscillation between around $t=6200$ and $t=7500$, where the target position is close to the ``pulled'' state of the plunger.
Solenoid itself has the mechanical characteristics that the plunger is rapidly `pulled'' at close position of ``pulled'' state, and it affects the PID control gains, and makes the position control more sensitive.

\section{Conclusion and Future Works}

In this paper, we described the algorithm to measure the stroke of solenoid using the electric characteristics of the solenoid, without mechanical attachments.
We also indicated that this measurement algorithm can be used with conjuncction of the stroke PWM control, and demonstrated the experimental results of controlling the solenoid stroke at intermediate position.

The precise measurement of the inductance and the positions the plunger, as well as the current for each position of the plunger, will be our future works to realize more accurate and stable mesurement and control of the solenoid stroke.
The precise evaluation of the calculated position of the plunger against the measured one are also reported in our future works.

The temperature dependency of the inductance and the resistance are also the parameters to consider, and will be discussed in our future works.

\begin{biography}
\profile{n}{Junichi Akita}{%
He received B.S., M.S. and Ph.D. degrees in electronics engineering from the University of Tokyo, Japan in 1993, 1995 and 1998 respectively. He joined the Department of Computer and Electrical Engineering, Kanazawa University as a research associate in 1998. He moved to the Department of Media Architecture, Future University Hakodate as an assistant professor in 2000. He moved to the Department of Information and Systems Engineering, Kanazawa University as an assistant professor in 2004. From 2022, he is a professor at School of Transdisplenary Science for Innovation, Kanazawa University. He is also interested in electronics systems including VLSI systems in the applications of human-machine interaction and human interface. He is a member of the Institute of Electronics, Information and Communication Engineers of Japan, Information Processing Society of Japan, and the Institute of Image Information and Television Engineering.}
\end{biography}

\end{document}